\documentclass[%
preprint,
]{elsarticle}

\usepackage{graphicx}
\usepackage{makecell}
\usepackage{subfig}
\usepackage{dcolumn}
\usepackage{natbib}
\usepackage{bm}
\usepackage{booktabs}
\usepackage{multirow}
\usepackage{siunitx}
\usepackage{enumitem}

\usepackage{color}
\usepackage[dvipsnames]{xcolor} 
\usepackage{colortbl}

\definecolor{DESYcyan}{RGB}{0,159,223}
\definecolor{DESYorange}{RGB}{241,143,31}
\definecolor{DESYred}{RGB}{235,90,45}
\definecolor{DESYdarkred}{RGB}{185,45,65}
\definecolor{DESYdarkblue}{RGB}{0,75,110}
\definecolor{DESYviolet}{RGB}{146,125,185}
\definecolor{DESYlila}{RGB}{80,80,155}
\definecolor{DESYblue2022}{RGB}{0,123,200}
\definecolor{DESYorange2022}{RGB}{235,110,15}

\usepackage{hyperref}
\hypersetup{
    colorlinks=true,
    linkcolor=DESYdarkblue,
    citecolor=DESYdarkred,
    filecolor=blue,      
    urlcolor=blue,
}

\usepackage{upgreek}
\usepackage{comment}
\usepackage{eurosym}

\makeatletter
\def\l@subsubsection#1#2{}
\makeatother

\begin{document}


\title{Neural Networks for ID Gap Orbit Distortion Compensation in PETRA III}
\author[inst1]{B.~Veglia}
\author[inst1]{I.~Agapov}
\author[inst1]{J.~Keil}

\affiliation[inst1]{organization={Deutsches Elektronen-Synchrotron DESY}, addressline={Notkestr. 85}, city={22607 Hamburg}, country={Germany}}

\date{\today}
\begin{abstract}

Undulators are used in storage rings to produce extremely brilliant synchrotron radiation. 
In the ideal case, a perfectly tuned undulator always has a first and second field integrals equal to zero. But, in practice, field integral changes during gap movements can never be avoided for real-life devices. As they significantly impact the circulating electron beam, there is the need to routinely compensate such effects. 
Deep Neural Networks can be used to predict the distortion in the closed orbit induced by the undulator gap variations on the circulating electron beam. In this contribution several state-of-the-art deep learning algorithms were trained on measurements from PETRA~III. The different architecture performances are then compared to identify the best model for the gap-induced distortion compensation. It is found that realistic data, with several gaps moving simultaneously to arbitrary gaps must be used to train the neural networks. The deep feed-forward neural network was found to be more effective than the recurrent and convolutional neural networks.
\end{abstract}

\maketitle

\section{INTRODUCTION}

\let\thefootnote\relax\footnotetext{Work supported by the ACCLAIM Innovationspool Project (HGF)}
The storage ring PETRA III at DESY \cite{petra} is being operated since 2009 and is one of the brightest synchrotron radiation sources worldwide. It serves a broad international multidisciplinary user community at currently 25 specialized beamlines. With a storage ring energy of \SI{6}{\GeV}, it delivers mainly hard to high-energy X-rays for versatile experiments in a very broad range of scientific fields. In this way, PETRA III provides the environment for specialized experiments to address many of the challenges of the twenty-first century in energy, life and health, earth and environment, mobility, and information technology. The circumference of \SI{2304}{\m} makes it the largest storage-ring-based source worldwide. Table \ref{tab:petraIII} summarizes the main operational parameters of the PETRA III storage ring.
\begin{table}
    \centering
    \begin{tabular}{ccc}
    \hline
    \multicolumn{1}{l}{\textbf{Parameter}} & \multicolumn{2}{c}{\textbf{Value}} \\
    \hline
    \multicolumn{1}{l}{Energy /GeV} & \multicolumn{2}{c}{6} \\
    \multicolumn{1}{l}{Circumference /m} & \multicolumn{2}{c}{2304} \\
    \hline
    \multicolumn{1}{l}{Total current /mA} & 120 & 100 \\
    \multicolumn{1}{l}{Number of bunches} & 480 & 40\\
    \multicolumn{1}{l}{Bunch Population /$10^{10}$} & 1.2 & 12\\
    \hline
    \multicolumn{1}{l}{Emittance} {(horz.) /nm rad} & \multicolumn{2}{c}{1.3}\\
     \multicolumn{1}{r}{(vert.) /nm rad} &  \multicolumn{2}{c}{0.01}\\
    \hline
    \multicolumn{1}{l}{Beam Size at 5 m Undulator (high $\beta$ Section)}{(horz.) /$\mu$m} & \multicolumn{2}{c}{141.5}\\
     \multicolumn{1}{r}{(vert.) /$\mu$m} &  \multicolumn{2}{c}{4.9}\\
    \hline
    \multicolumn{1}{l}{Beam Size at 5 m Undulator (low $\beta$ Section)}{(horz.) /$\mu$m} & \multicolumn{2}{c}{34.6}\\
     \multicolumn{1}{r}{(vert.) /$\mu$m} &  \multicolumn{2}{c}{6.3}\\
    \hline
    \multicolumn{1}{l}{Beam Size at 10 m Undulator}{(horz.) /$\mu$m} & \multicolumn{2}{c}{141.6}\\
     \multicolumn{1}{r}{(vert.) /$\mu$m} &  \multicolumn{2}{c}{6.6}\\
    \hline
    \end{tabular}
    \caption{PETRA III electron beam parameters.}
    \label{tab:petraIII}
\end{table}

Undulators are the most powerful generators of synchrotron radiation at storage rings, they are a type of insertion devices (or IDs).
The angle-dependent emitted wavelength can be calculated as:

\begin{equation}
    \lambda(\theta) \approx \frac{\lambda_U}{2 \gamma^2}\left(1 + \frac{K^2}{2}+\gamma^2\theta^2\right)
\end{equation}

Where $\lambda_U$ is the period length of the undulator, $\gamma$ is the Lorentz factor, $\theta$ is the emission angle and dimensionless quality K is the deflection (or undulator) parameter that can expressed as:

\begin{equation}
    K = \frac{e}{2\pi\, m_e\, c}B_0\,\lambda_U = 0.9336\,B_0[T]\,\lambda_U[cm]
\end{equation}
with $e$ the electron charge and $m_e$ its mass, $c$ the speed of light and $B_0$ the peak magnetic field on the undulator axis. 

\begin{figure*}[!tbh]
    \centering
    \includegraphics[width=\textwidth]{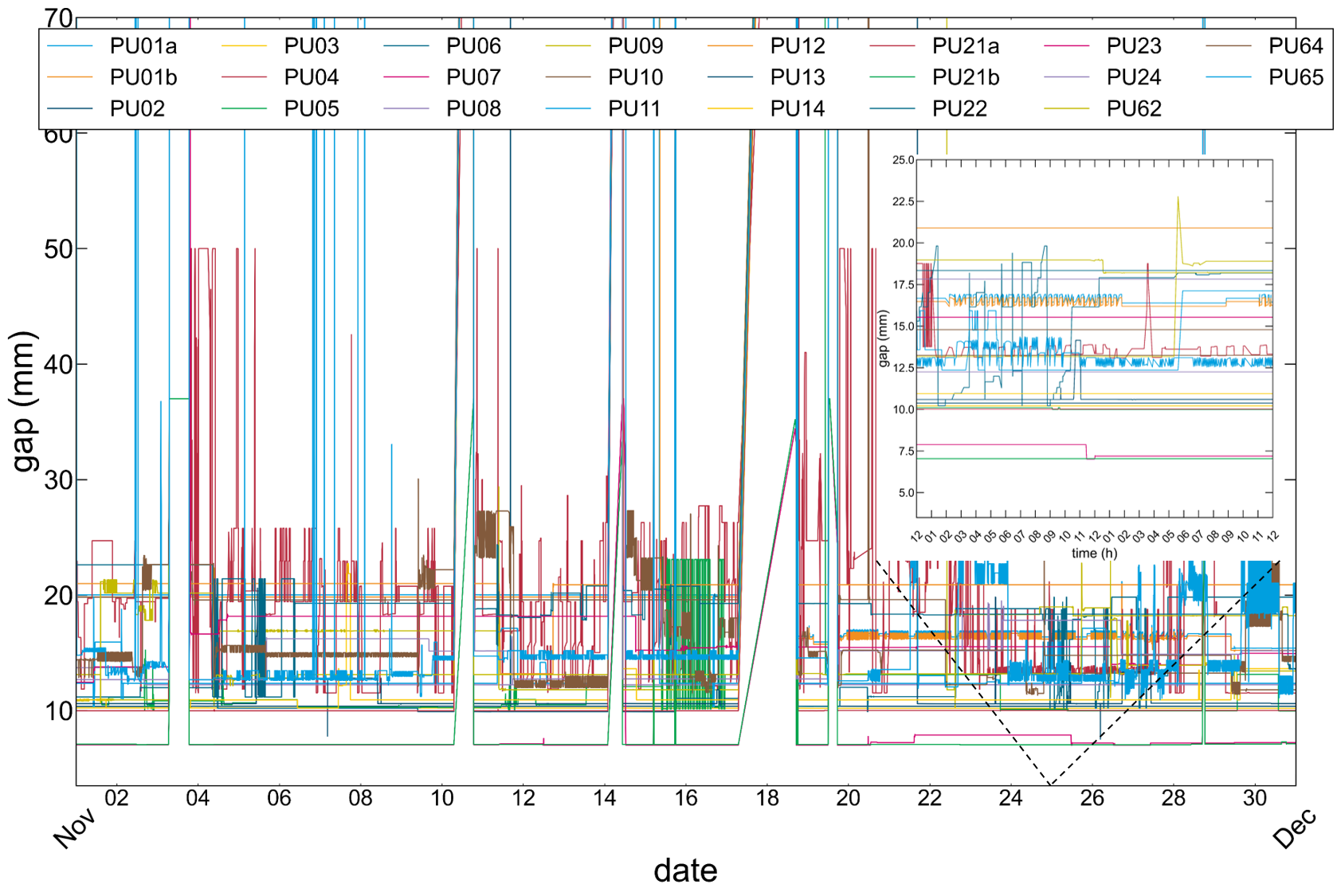}
    \caption{Recorded gap sizes of PETRA III IDs over a month period at the end of 2021.}
    \label{fig:gap_var1}
\end{figure*}

\begin{figure}[!tbh]
    \centering
    \includegraphics[width=0.65\columnwidth]{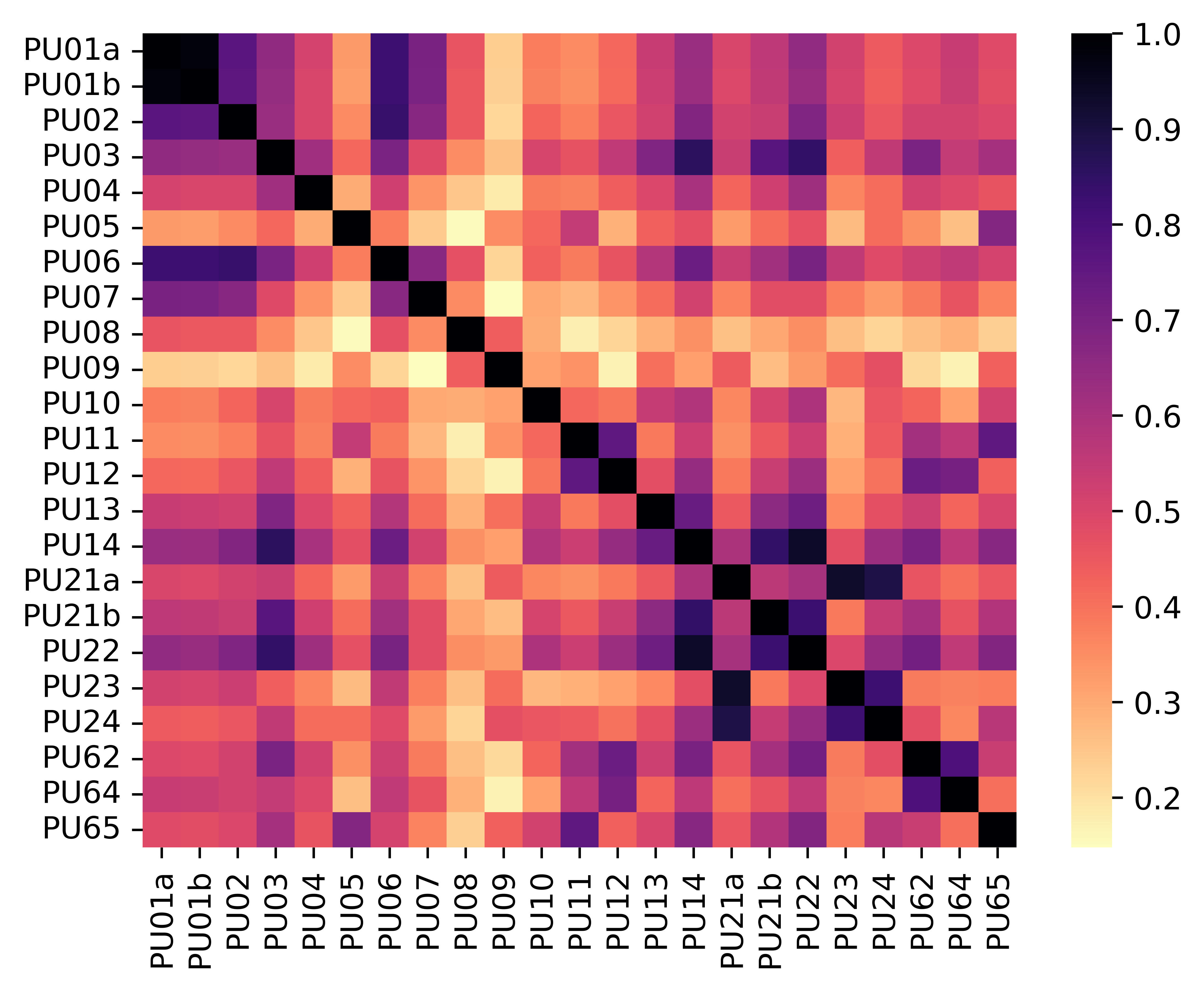}
    \caption{Heatmap of the correlation matrix of gap movements over the year period.}
    \label{fig:corr}
\end{figure}

Radiation wavelength is adjusted by changing the undulator gap $g$, that is, the distance of the opposing magnets, as shown in the following relation

\begin{equation}
    K \approx 0.168\, B_r\, \lambda_U \exp{\left( - \pi \, \text{g}/ \lambda_U \right)}
\end{equation}

where $B_r$ is the on-axis magnetic field.
Figure~\ref{fig:gap_var1} shows that the changes applied by users are very frequent, do not follow specific patterns and, as seen in Fig.~\ref{fig:corr} they are in general statistically uncorrelated (except few cases, P01a and P01b for example are two \SI{5}{\m} long undulators synchronized and considered as one \SI{10}{\m} long device). During operations normally most of the IDs would be moving to small gaps and very rarely be fully open at the same time. Table \ref{tab:beamlines} lists the beamlines operating with undulator radiation. Some experiments perform wide scans of wavelengths while others apply small fast variations. The beamline users have hence the control over the undulator movements, while the accelerator operation is in charge of compensating any effect that those might induce on the circulating electron beam.

\begin{table*}
    \centering
    \renewcommand{\cellset}{\renewcommand{\arraystretch}{0.8}}
    \begin{tabular}{ccccc}
    \hline
        \textbf{Beamline} & \textbf{Length} & \textbf{Period length $\lambda_U$}& \textbf{Gap range}  & \textbf{Energy range} \\
        & (m) & (mm) & (mm) & (keV) \\
        \hline
        P01 a and b & 2 x 5 & 31.4 & 12.7 - 220  & 2.5 - 80 \\
        \midrule
        \makecell{P02.1\\ P02.2}& 2 & 23.0 & 9.9 - 220 & \makecell{60\\ 25.7, 42.9, 60}\\
        \hline
        P03 & 2 & 29.0  & 10.2 - 217 & 9 - 23 \\
        \hline
        P04 &  4.9 & 65.6 (APPLE*)& 11.5 - 148 &  0.25 – 3.0 \\
        \hline
        P05 & 2 & 29.0 & 9.9 - 220 & 5 – 50  \\
        P06 & 2 & 31.4 & 9.8 - 220 & 5 - 45 \\
        P07 &  4 & 21.2 (In-Vacuum) & 7.0 - 40  &  30 – 200  \\
        \hline
        P08  & 2 & 29.0 & 9.9 - 218   & 5.4 – 29.4  \\
        P09  & 2 & 31.4 & 9.8 - 220  & 2.7 -31 \\
        \hline 
        P10  & 5 & 31.4 & 9.7 - 220  & 5 -10 \\
        \hline
        P11 &  2 & 31.4 & 9.9 - 220 & 5.5 – 30 \\
        P12 & 2 & 21.2 & 9.9 - 220  & 4 – 20 \\
        \hline
        P13 &  2 & 29.0 & 9.8 - 220  & 4.5 – 17.5 \\
        P14 & 2 & 21.2 & 9.8 - 220  & 7 – 26.7 \\
        \hline
        P21a & 2 & 29.0 & 9.9 - 220 & 52, 85, 100\\
        P21b & 4 & 21.2 (In-Vacuum) & 7.0 - 40 & 40 - 150 \\
        \hline
        P22 & 2 & 32.8 & 9.6 - 220  & 2.4 - 30 \\
        P23 & 2 & 31.4 & 9.9 - 220  & 5 - 35 \\
        \hline
        P24 & 2 & 21.2 & 9.8 - 220  & 8, 17 - 44 \\
        P25 & & & in preparation\\
        \hline
        P62 & 2 & 31.4 & 9.9 - 220  & 3.5 - 35 \\
        P63 & & & in preparation\\
        \hline 
        P64 & 2 & 32.8 & 10.4 - 220  & 4 - 44 \\
        P65 & 0.4 & 32.8 & 10.5 - 220 & 4 - 44 \\
        \hline

    \end{tabular}
    \caption{PETRA III beamlines \cite{Bieler_2022}. *APPLE is an abbreviation for Advanced Planar Polarized Light Emitting undulator, a device made of four independent arrays of permanent magnets which can be adjusted to control the polarisation of
    the radiation emitted.}
    \label{tab:beamlines}
\end{table*}

In 3rd generation light sources the orbit must be kept stable to within a few percent of the electron beam size (see Tab.\ref{tab:petraIII}). The field changes due to gap variations thus need to be compensated to not disrupt other users. 
With the advent of 4th generation storage rings delivering high-brightness x-ray beams with high coherent flux, electron beam sizes will become smaller. This is the case of the upcoming upgrade PETRA IV \cite{petraiv}, which will also present an increased number of IDs, making the orbit stability even more challenging and important.

The magnetic fields of IDs introduce perturbations to the circulating electron beam and hence affect the linear and nonlinear beam dynamics of the electron beam in the storage ring.
The primary effect of the undulator on the electron beam trajectory is determined by the field integrals \cite{undul}. 

In the ideal case, a perfectly tuned undulator with an antisymmetric magnet structure always has a first field integral equal to zero. But due to imperfections, magnetic field degradation caused by radiation damage \cite{ID_radiation2,ID_radiation} as well as concentration of ambient magnetic fields by undulator poles, field integral changes during gap movements can never be avoided for real-life devices. The traditional approach to compensate this effect is based on measurements of the response to gap movements on the machine orbit (performed only once during commissioning) to calculate look-up tables \cite{Karabekyan}. At PETRA such tables were empirically determined by performing measurements (closing one ID at the time) during commissioning with the fast orbit feedback being disabled. From these data a look-up table for each undulator control system is calculated in order to power the compensation coils according to the current gap value. 
\begin{figure*}[t!]
        {\centering
        \textbf{PU08}\par\vspace*{-2mm}}
        \subfloat[]{%
            \includegraphics[width=.48\linewidth,trim={0 0 0 0.55cm},clip]{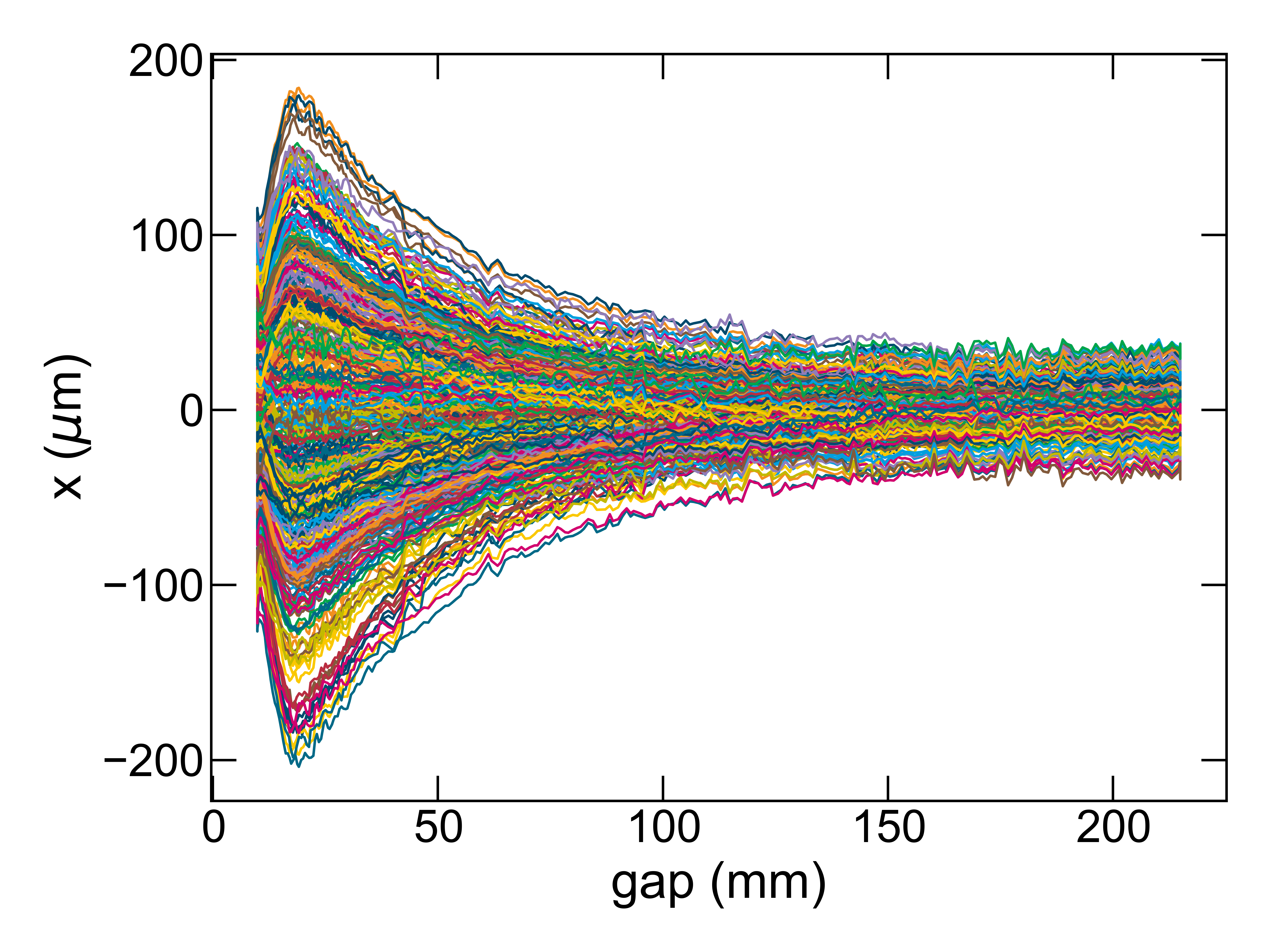}%
            \label{subfig:a}%
        }\hfill
        \subfloat[]{%
            \includegraphics[width=.48\linewidth,trim={0 0 0 0.55cm},clip]{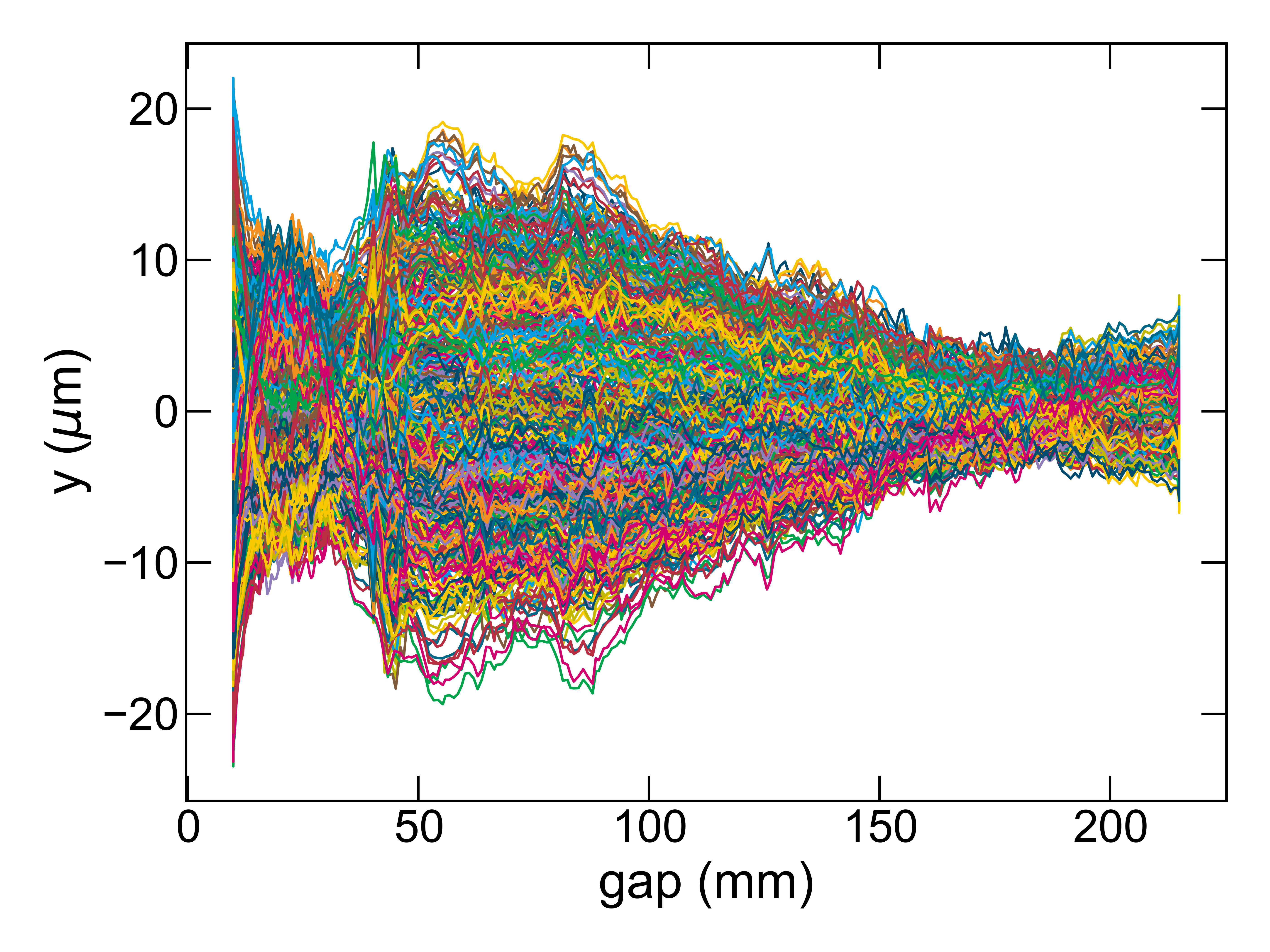}%
            \label{subfig:b}%
        }\\
        \par
        {\centering
        \textbf{PU14}\par\vspace*{-2mm}}
        \subfloat[]{%
            \includegraphics[width=.48\linewidth,trim={0 0 0 0.55cm},clip]{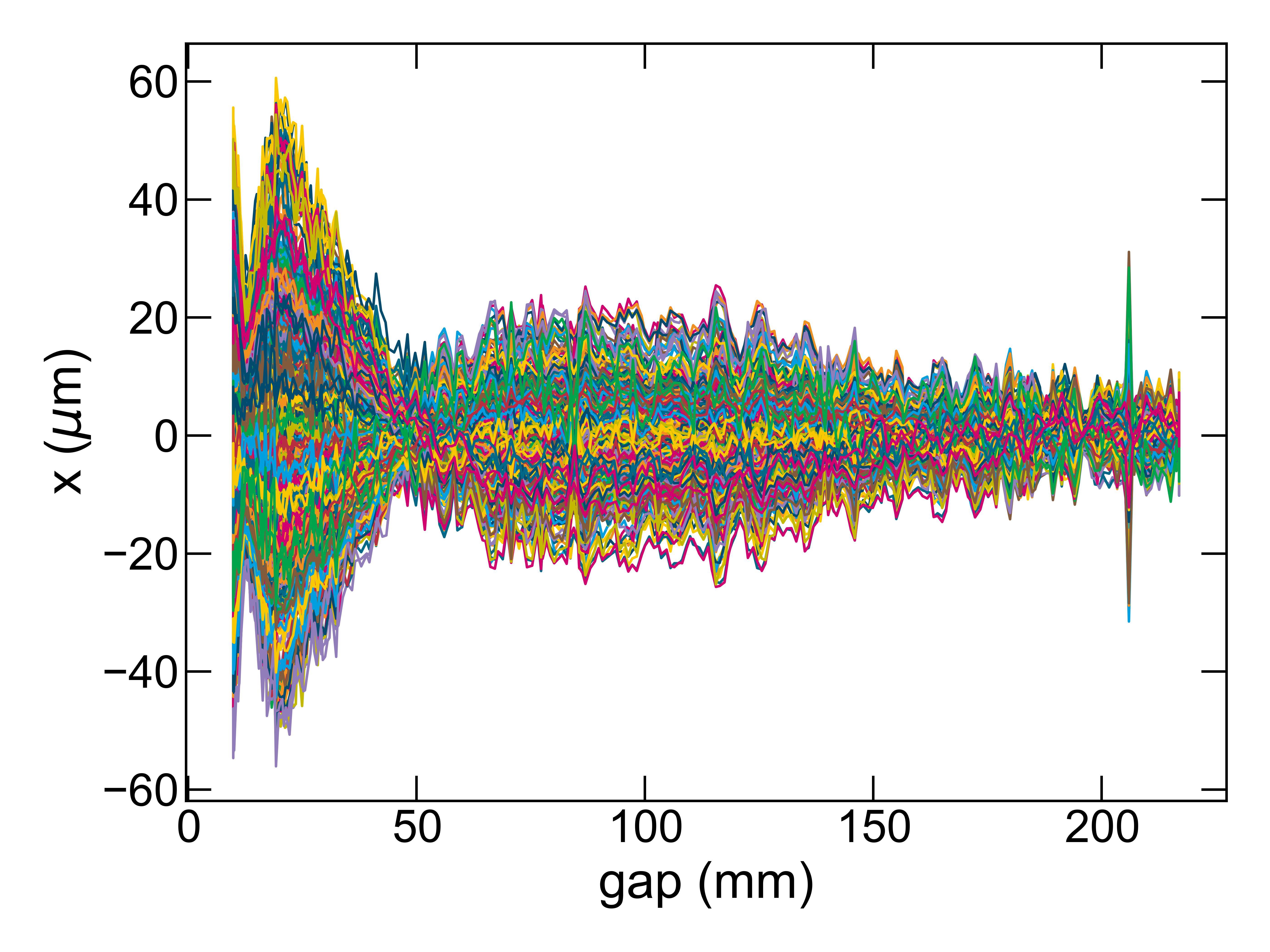}%
            \label{subfig:c}%
        }\hfill
        \subfloat[]{%
            \includegraphics[width=0.48\linewidth,trim={0 0 0 0.55cm},clip]{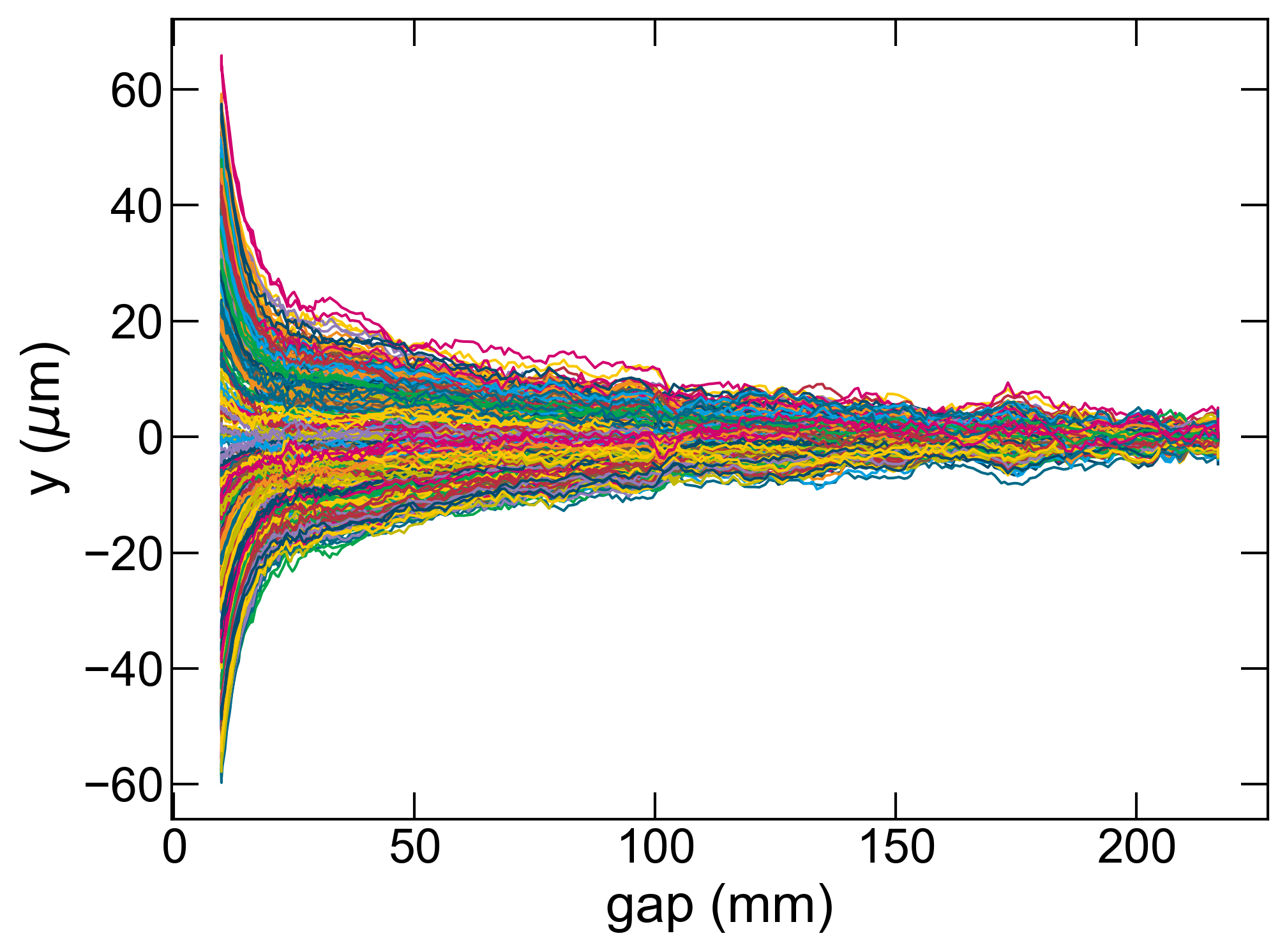}%
            \label{}%
        }
        \caption{Measurement of two IDs varying gap vs. the horizontal (a and c) and vertical (b and d) orbit. Each coloured curve corresponds to a different BPM.}
        \label{fig:4pus}
    \end{figure*} 
 Measurements of the individual impact of the IDs gap variations on the closed orbit are shown in Fig.~\ref{fig:4pus}. Each plotted line represents the detected position at the corresponding BPM. 

The ability of neural networks to learn non-linear patterns \cite{CARPENTER}, approximating any complex function through an ensemble of latent variables, makes them ideal candidates to capture and model the effect of undulator movements on the circulating electron beam, which presents complex nonlinear behaviour. 
This contribution shows how the application of a model-independent neural network could allow for such closed orbit distortion corrections. 

\section{NEURAL NETWORKS}
Neural networks (NNs) have proved to be most effective for nonlinear function fitting, both theoretically and empirically \cite{hornik,AGC20}. Recently, improved techniques from the fields of machine learning (ML) and artificial intelligence (AI) have been incorporated into the design of control systems for particle accelerators. In particular, techniques based on neural networks (NNs) are well-suited to modeling, control, and diagnostic analysis of complex, time-varying systems, and systems with large parameter spaces \cite{Edelen, Ivanov}.

\subsection{Architectures considered}
\begin{figure}[!htb]
    \centering
    \includegraphics[width=0.9\columnwidth]{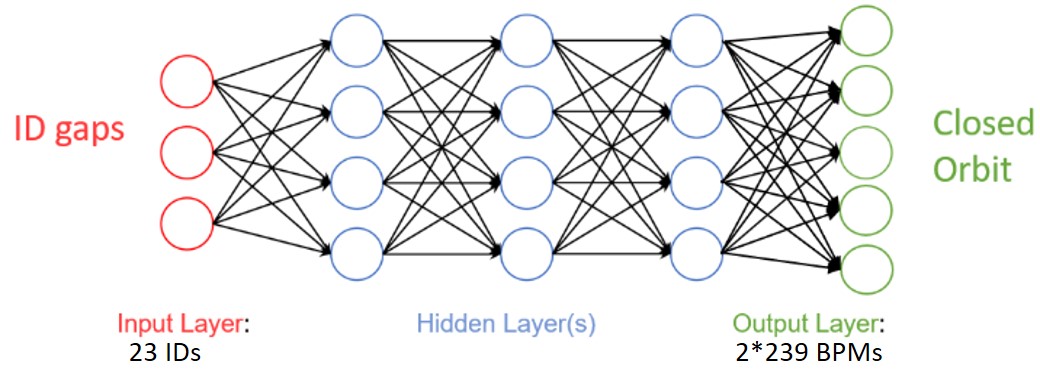}
    \caption{Schematic layout of a Neural Network mapping the ID gaps to the closed orbit.}
    \label{fig:NN}
\end{figure}

Here, we propose a NN approach which, taking arbitrary ID gap configurations as input, predicts the corresponding impact on the electron beam orbit giving the transverse position at the BPMs as output. The NN are capable to learn complex nonlinear relationships between the ID settings and transverse position using training data.  Figure~\ref{fig:NN} illustrates a schematic representation of a generalized NN architecture mapping the ID gap sizes to the closed orbit.

Each training input is loaded into the neural network in a process called forward propagation. Once the model has produced an output, this predicted output is compared against the given target output in a process called back-propagation. Stochastic gradient descent is used to minimize a loss function, quantifying the error produced by the model. The parameters of the model are updated at each iteration (or epoch) so that its output better fits the target.

The NNs are implemented using Keras with Tensorflow backend \cite{keras,tensorflow}, using mean squared error as the loss function. The models are trained using the back-propagation method employing the Adam optimizer \cite{Adam} for 200 or fewer epochs.  The training takes between 10 and 50 minutes on a single desktop-class CPU depending on the model. A variety of NN architecture features was screened including regularization methods and activation functions in order to optimize the model structures, described in Sect.~\ref{sect:sweep}. Weights\&Biases \cite{wandb} was used to automate hyperparameters tuning, exploring the space of possible models and allowing to identify the best suited ones for the problem. In the following subsections the model architectures considered are described.

\subsubsection{FFNN}
The simplest architecture is a NN where the output from the "neurons" of one layer is used as input to the next layer. Such networks are called feed-forward neural networks (FFNN) or Multilayer Perceptron (MLP) and can use linear or non-linear activation functions. The information is always propagated forward within the network and there are no loops. In this study we considered a FFNN with only one hidden layer (shallow) and a deeper one with three hidden layers. We expect that the deep FFNN achieves higher accuracy than the shallow one (at the same computational power) thanks to its ability to better extract features.

\subsubsection{RNN}
\begin{figure}[!tbh]
    \centering
    \includegraphics*[width=0.8\columnwidth]{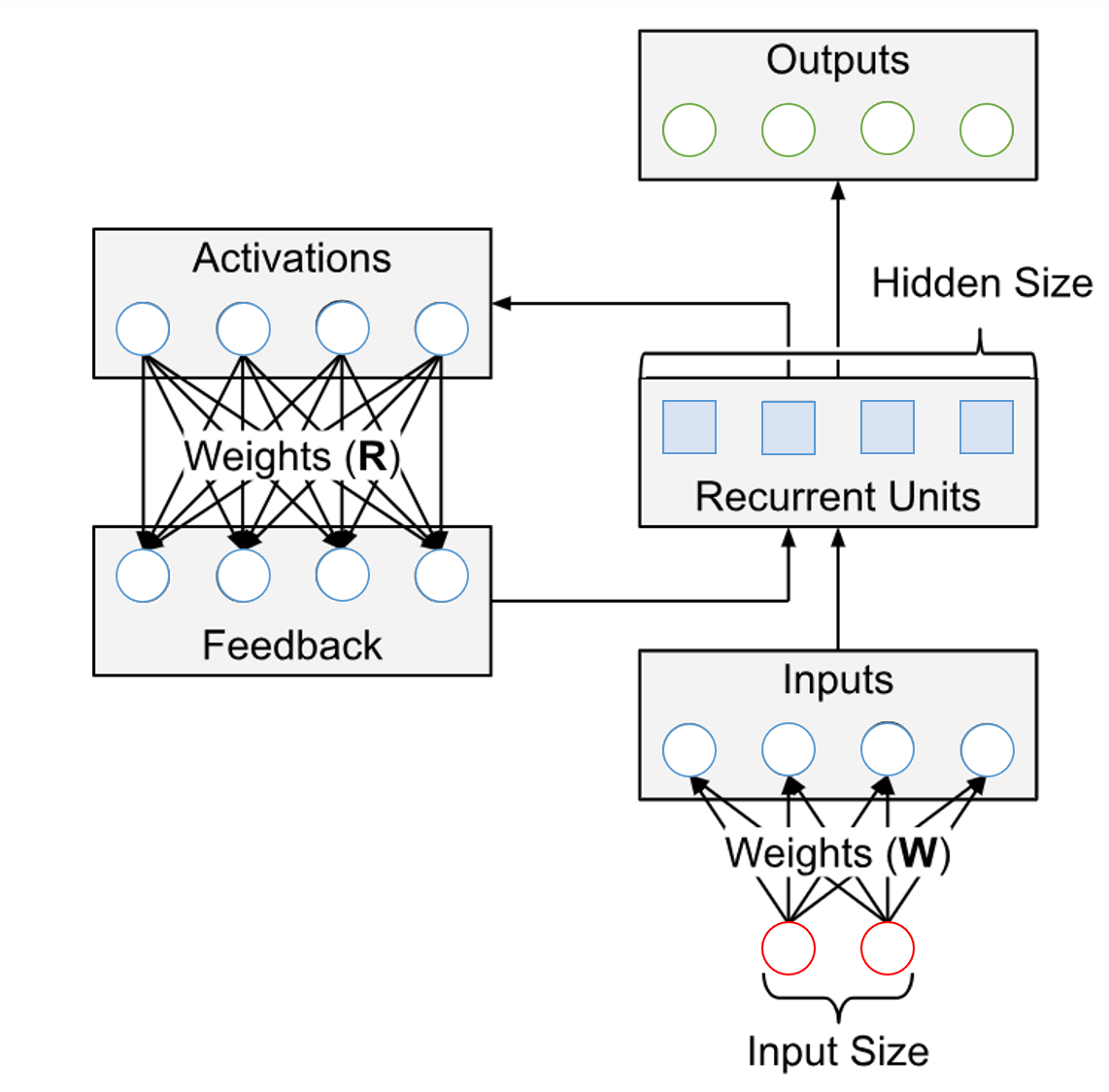}
    \caption{Schematic representation of a single recurrent layer with two inputs and four hidden units.}
    \label{fig:RNN}
\end{figure}

Recurring neural networks, or RNNs, rely on a simple principle, feeding the output of a layer back to itself, thus creating a feedback loop between different layers of the network, allowing output from some neurons to affect subsequent input to the same neurons as represented in Fig.~\ref{fig:RNN}. This enables it to exhibit temporal dynamic behavior acting like a neural memory \cite{elman}. Working with sequential data, the order in which information appears is important and by feeding back the previous input to the current layer, it allows to apply updates in the context of the sequence seen so far. The recurrent network keeps a record of the input signal finding patterns between different parts of the input. The specific architecture studied in this contribution consisted of two hidden layers, each composed of recurrent units, followed by the fully connected output layer. 

\subsubsection{CNN}
Convolutional layers represent major building blocks in many deep neural networks. The design was inspired by the visual cortex, where individual neurons respond to a restricted region of the visual field known as the receptive field. 
Convolutional neural networks (CNNs) \cite{Lecun} get their name because their layers use convolution to learn a feature map from the input data. Convolution is a mathematical function that is used in place of matrix multiplication in the layers of CNN. There are four main components in CNNs: (a) convolution filters, (b) nonlinear activations, (c) spatial coarsening (via pooling or strided convolution), (d) a prediction module, often consisting of fully connected layers that operate on a global instance representation.
The NN used for the closed orbit prediction consist of a 1D convolutional layer, followed by the application of a 1D pooling and two fully connected layers  \cite{1Dcnn}. 

\section{MEASUREMENTS}
\subsection{Training data}
The data used for training the neural networks was acquired in July 2022 during a dedicated machine study shift by scanning 18 IDs gaps within their full range and recording the orbit at each of the functioning 239 BPMs. The values for the remaining undulators were taken from previous measurement campaigns performed following the same procedures.
To observe the raw impact of the ID gap variation on the circulating electron beam, the fast orbit feedback system was disabled. The previously mentioned Figure~\ref{fig:4pus} shows the measurements of the orbit varying with the gap opening in one of the IDs while the other were fully open. For each measurement the IDs were moved towards the minimum gap in steps of 0.2 to \SI{1}{\milli\metre} (depending on the undulator type) every half second and then reopened again and kept to the maximum gap during the movement of the other IDs. The maximum and minimum gap size depends on the type of undulator, the so-called 'in-vacuum' ones, for example, reach very small gaps but are constrained by the vacuum chamber when opening. The number of data points collected for each ID goes between 325 (in-vacuum) to 704 (the APPLE II has a slower gap movement speed in respect to other IDs). To train the network the collected dataset was augmented by linear interpolation extending the dataset to 704 data points for each ID. 

Models for each of the four different architectures (Shallow and Deep FFNN, RNN and CNN) were trained on 80\% of the training measurements. The remaining 20\% of the data was used for validation in order to gauge the models' ability to predict the electron beam orbit.

\subsection{Hyperparameter Sweeps}\label{sect:sweep}
\begin{figure}[!tbh]
    \centering
    \includegraphics[width=0.8\columnwidth]{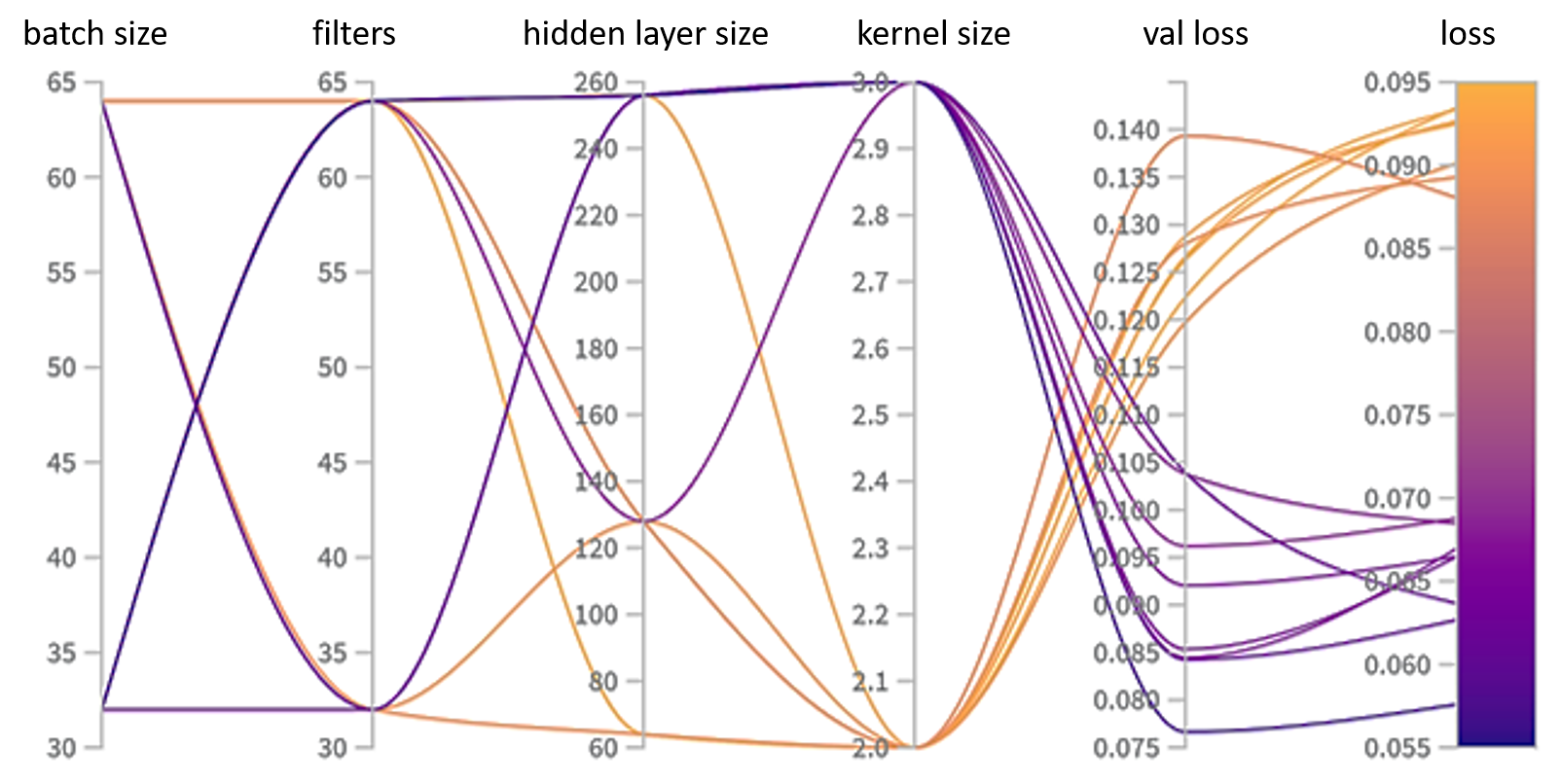}
    \caption{Hyperparameter tuning for the CNN.}
    \label{fig:Sweep3}
\end{figure}

Hyperparameters, such as the learning rate and the number of hidden layers, play a crucial role in determining a machine learning model's performance.
A small difference in a single hyperparameter can lead to a large performance variation. Therefore, a central component of machine learning research is to find the correct set of hyperparameters for a given task. It can be challenging to identify optimal hyperparameters for a data distribution sequentially since the parameter search space is large and has very few optimal values. Systematic hyperparameter sweeps were performed with Weights\&Biases in order to find the best working combination for each of the NN structures considered, allowing to fairly compare their performance. The goal of the optimization is to minimize the loss. The following hyperparameters were investigated:
\begin{description}
    \item[Activation function] determines the range of values of activation of an artificial neuron. This is applied to the sum of the weighted input data of the neuron. Activation functions introduce non-linearities, allowing neural networks to learn highly complex mappings between inputs and outputs. 
    \item[Batch size] defines the number of samples propagated through the network.
    \item[Hidden layer size] is the number of neurons in a layer which controls the representational capacity of the network, at least at that point in the topology.
    \item[Learning rate] controls the step size for a model to reach the minimum loss function \cite{NNbook}.
    \item[Dropout] is a regularization technique to avoid overfitting by randomly dropping out neurons during training \cite{dropout}.
    \item[Filters] indicates the amount of filters applied to the input in the convolutional layer.
    \item[Kernel size] specifies the amplitude of the convolution window in convolutional layers.  
\end{description}
\begin{figure}[!tbh]
    \centering
    \includegraphics[width=0.9\columnwidth]{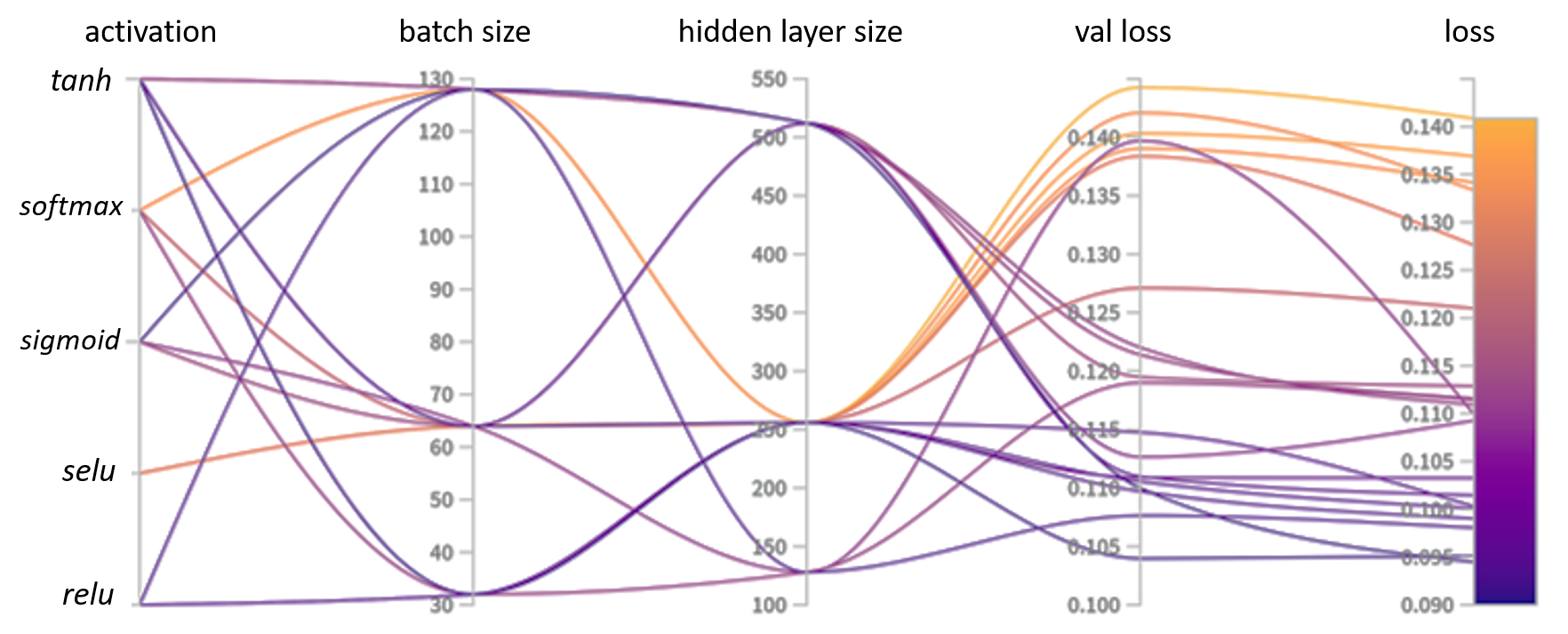}
    \caption{Representation of the optimization of various hyperparameters for the shallow FFNN. The optimization aims at minimizing the loss. In the diagram the validation loss ("val loss" in figure) is also displayed for reference. }
    \label{fig:FFNN1}
\end{figure}
\begin{figure}[!tbh]
    \centering
    \includegraphics[width=0.8\columnwidth]{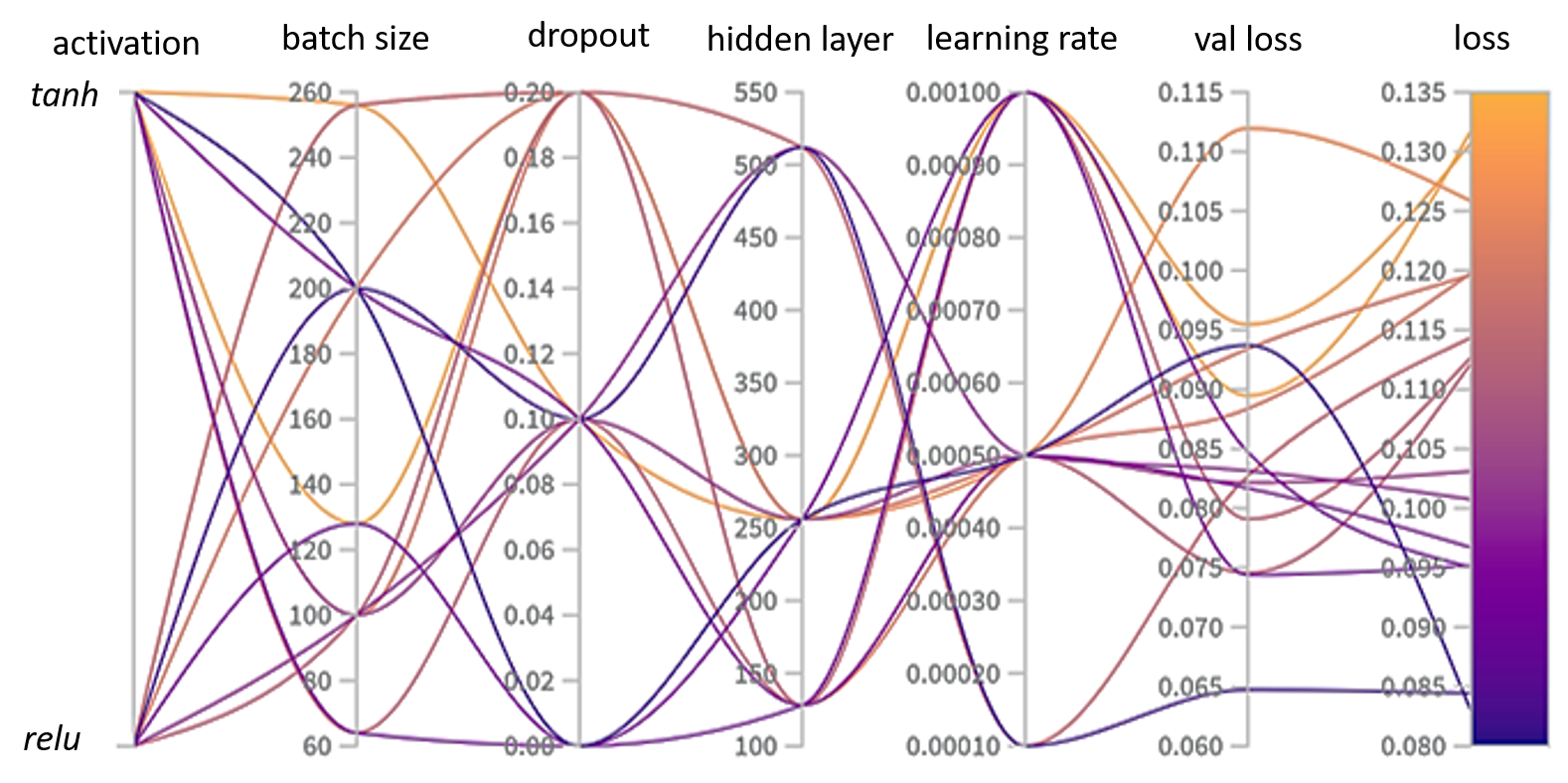}
    \caption{Graphical representation of the hyperparameter tuning for the deep FFNN.}
    \label{fig:Sweep}
\end{figure}
\begin{figure}[!tbh]
    \centering
    \includegraphics[width=0.9\columnwidth]{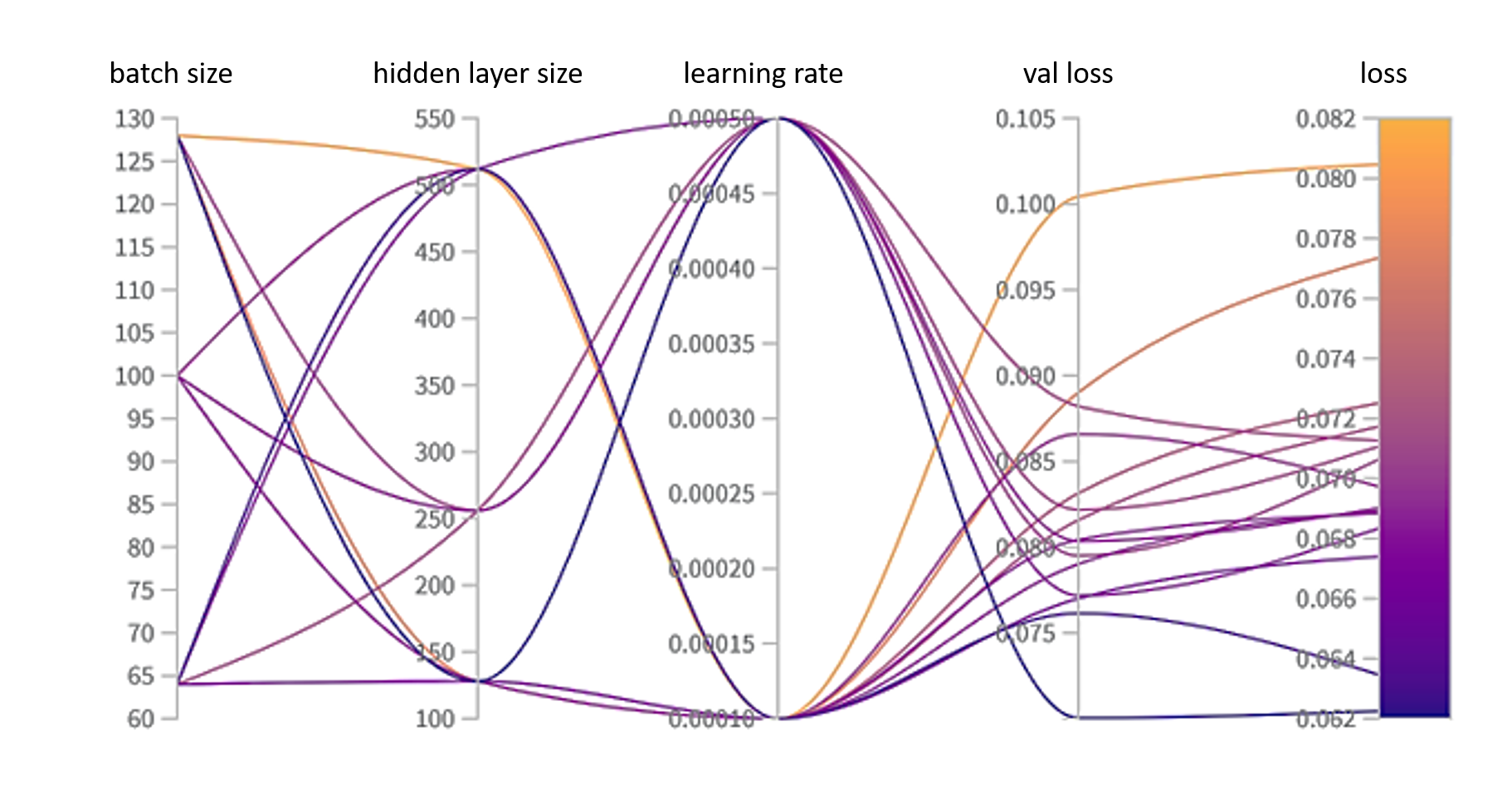}
    \caption{Graphical representation of the hyperparameter tuning for the RNN.}
    \label{fig:Sweep2}
\end{figure}
In Figures~\ref{fig:FFNN1} to \ref{fig:Sweep3} the impact of the hyperparameters relevant for each architecture on the training loss and validation loss  is displayed. Each vertical axis represents one hyperparameter while the colorbar indicates the loss relative to each hyperparameter combination tested. 

\begin{figure}[!tbh]
    \centering
    \includegraphics[width=0.73\columnwidth]{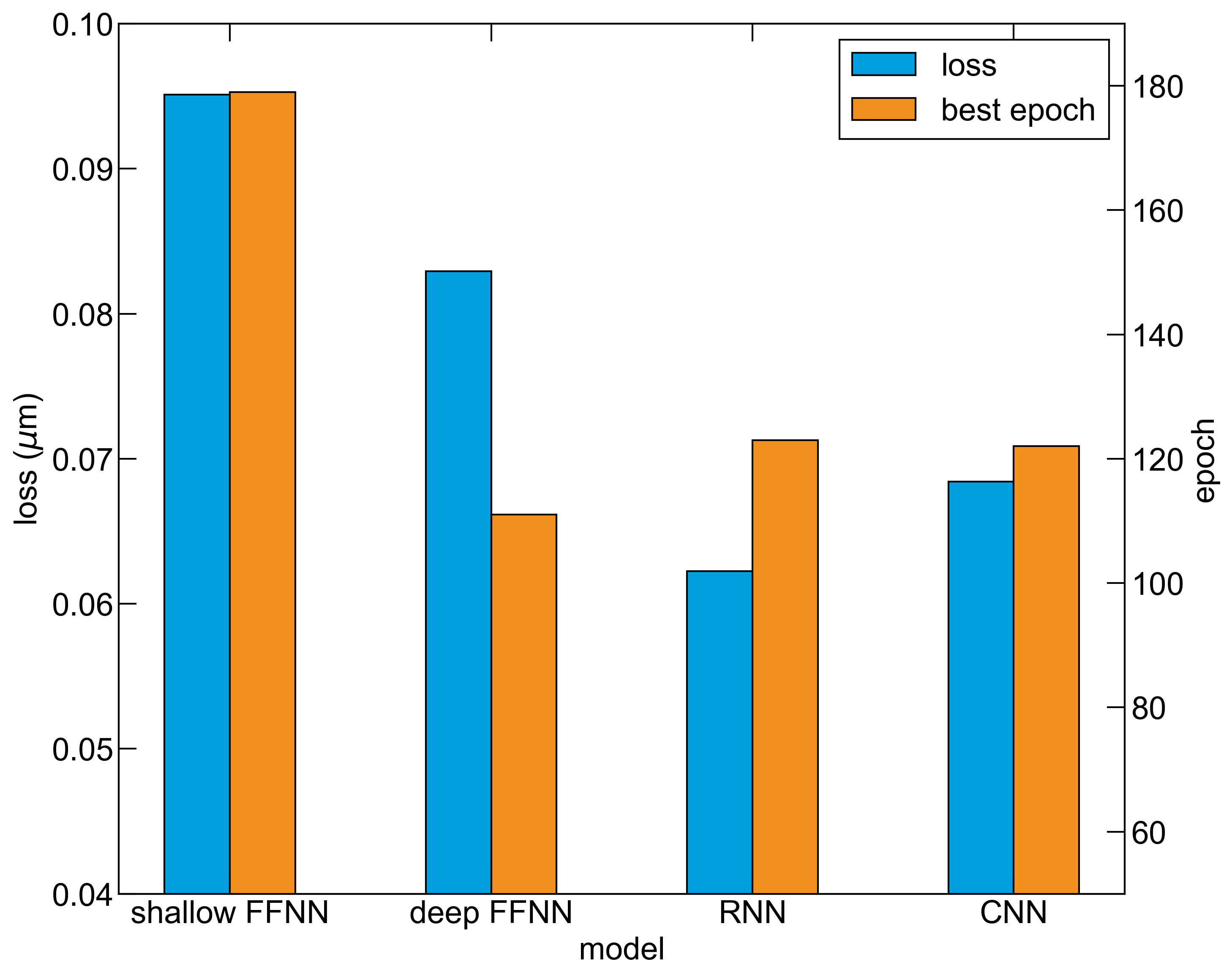}
    \caption{Comparison of the performance of the four different model architectures.}
    \label{fig:comp}
\end{figure}

The training results were compared for each architecture once the hyperparameters were optimized. Figure \ref{fig:comp} shows the final training loss and the epoch at which the training stopped. The loss value gives an indication on how well the model fits the data, while the number of epochs specify how many training iterations were required. 

\begin{figure}[!tbh]
    \centering
    \includegraphics[width=0.9\columnwidth]{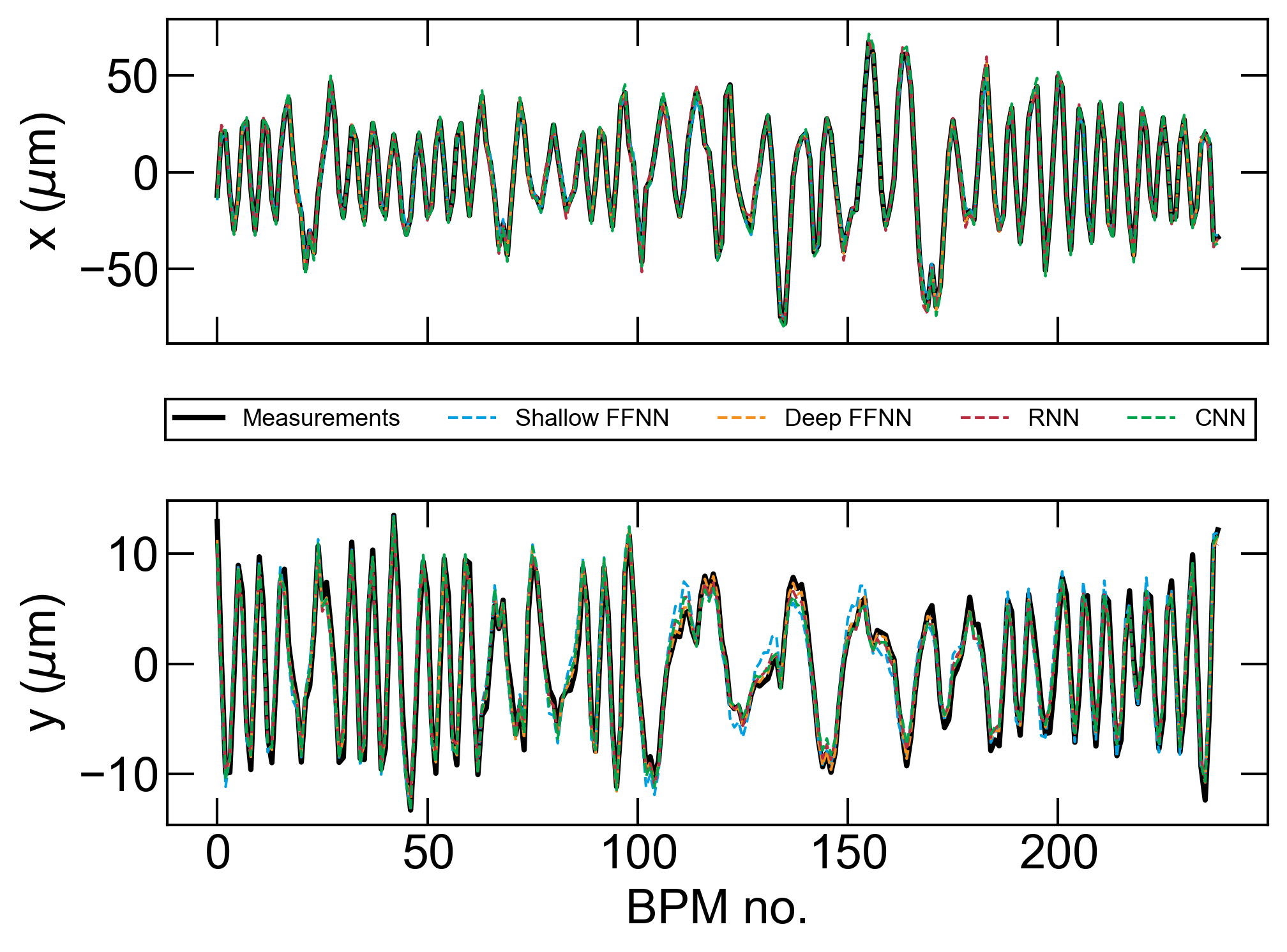}
    \caption{Horizontal (top) and vertical (bottom) beam orbit predictions by the trained NNs (dotted coloured lines) vs validation values of the training dataset (black solid line).}
    \label{fig:pred}
\end{figure}

The convolutional and recurrent structures outperform the fully connected NNs reaching better accuracy in a reasonable amount of epochs. This is due to the format of the measurements used for training, where the gap size was recursively increased and decreased for each undulators. The input data exhibited a serial structure that is better modelled by NNs containing recurrent or convolutional features. 
As seen in Fig.~\ref{fig:pred} the tuned NNs, although giving slightly different outputs, are generally able to accurately model the training data.

\subsection{Multi-ID data}
\begin{figure}[!tbh]
    \centering
    \includegraphics[width=\textwidth]{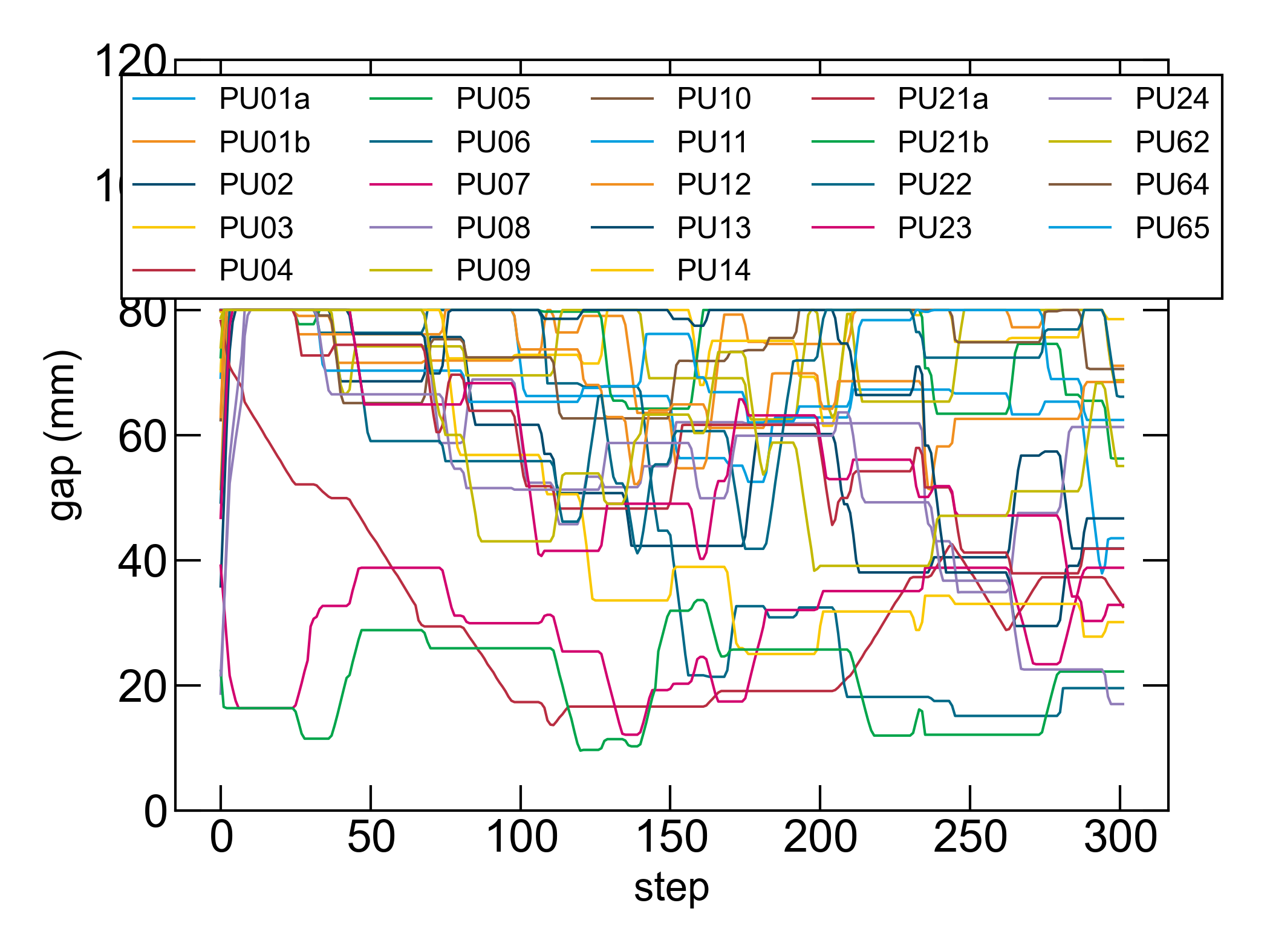}
    \caption{Example of the random gap variations used for the operational data measurements. Each colour represents a PETRA III Undulator (PU).}
    \label{fig:gap_var}
\end{figure}

Since the training set was measured moving only one undulator at the time, to test the trained network predictions with realistic operative gap values orbit, "multi-ID" data was collected. In this set multiple undulators were closed at the same time, observing their combined impact on the orbit. The undulators were moved to randomly chosen gaps as shown in Fig.~\ref{fig:gap_var}.

Constrains were in place to avoid losing the beam by inducing a too large transverse kick. At each step a random gap value (taken form a univariate normal distribution of mean \SI{0}{\mm} and variance \SI{10}{mm}) is assigned to a random subset of the insertion devices. 
\begin{figure}[!tbh]
    \centering
    \includegraphics[width=\textwidth]{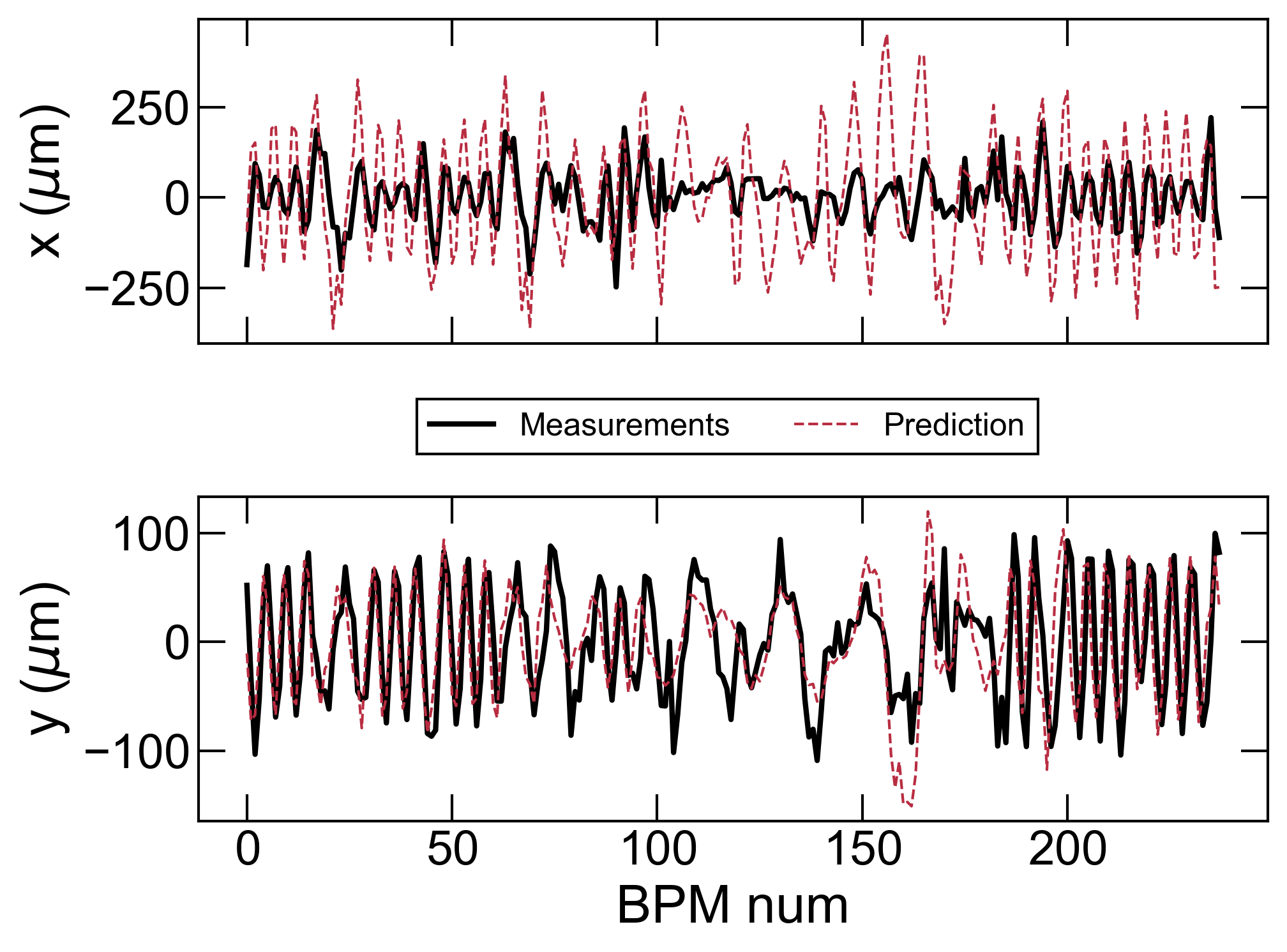}
    \caption{Measured orbit distortion induced by multiple IDs closing compared to the prediction obtained with the RNN model trained on the initial data only.}
    \label{fig:diff}
\end{figure}

While the IDs were moving towards the desired position, the orbit was recorded every second. Three dedicated measurement campaigns were performed on different dates, collecting a total of 1940 data points. 
Comparing the expected orbit distortions obtained with any of the trained and optimized model, it was found that they significantly differ from the measurements (rms error of  \SI{81}{\um} in the horizontal plane and  \SI{15}{\um} in the vertical). Fig.~\ref{fig:diff} shows the comparison of the closed orbit measured with an operational IDs configuration and the prediction calculated with the RNN model.

\begin{figure}[!tbh]
    \centering
    \includegraphics[width=0.9\columnwidth]{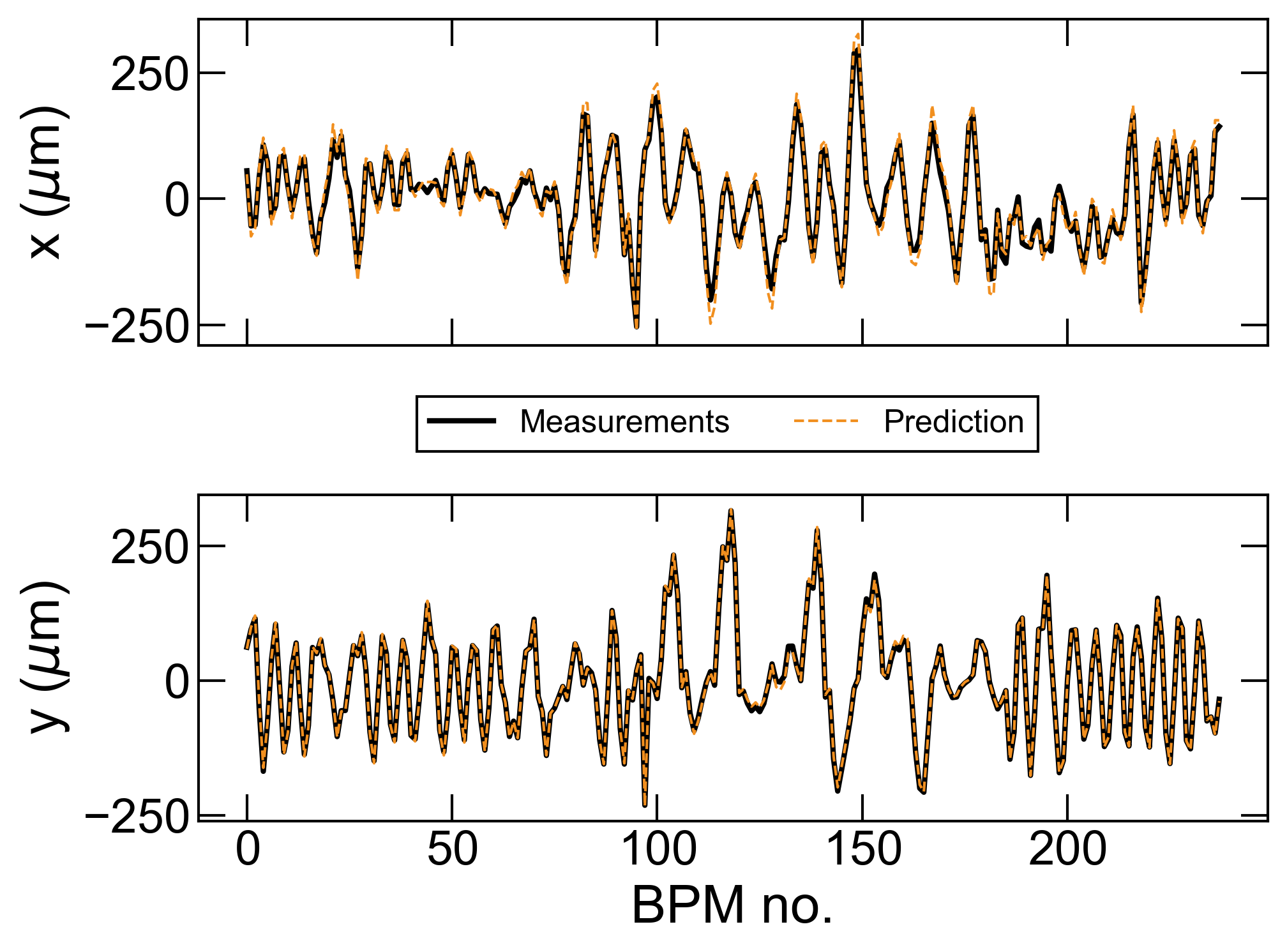}
    \caption{Horizontal (top) and vertical (bottom) beam orbit predictions by the re-trained deep FFNN (dotted yellow lines) vs. the measured value (black solid line). The IDs gap configuration was randomly selected within a test subset of the multi-ID data.}
    \label{fig:pred2}
\end{figure}

The RNN, which achieved the smallest loss on training data was thus retrained on a subset of the new data (randomly distributed between training, validation and test sets). The updated model was still not able to predict the orbit distortion. It was the deep fully connected architecture, trained on the whole dataset including the new multi-ID measurements, that could effectively model the IDs effects as shown in Fig.~\ref{fig:pred2}. The predictions are made on a test subset of data not used for training. 
Additionally, we introduced another regularization method: early stopping. By monitoring the validation loss during training and halting the process when a minimum is reached, overfitting can be effectively prevented, ensuring that the model performs well on unseen data. The rms error on $x$ is \SI{2}{\um} while on $y$ is \SI{0.89}{\um}.

\section{DISCUSSION}
The NNs take as input a vector containing the gap size of each ID for any operational configuration and give as output the predicted transverse position at the BPM locations.  

With the final model it is possible to scan 23 IDs through their entire operational parameter space and evaluate the expected transverse position at the BPMs. A FFNN was successfully trained to make accurate predictions of the beam transverse position.

Initially, it was expected that a network effectively trained on measurements taken moving one ID at the time, would be able to predict the orbit distortion caused by multiple IDs moving simultaneously. Since such prediction were inaccurate, multi-ID data were introduced into the training. The obtained model was then tested to better understand the relation between the undulator movements and the electron orbit response. In order to verify whether linear combinations of the explanatory variables (ID gaps) can be used to predict the response variable (closed orbit distortion) we created a linearity test that is defined by the following.

Having $F_j$ as the orbit at BPM $j$ and 

$$g_l = (0,0,0,0,0,x_l,0,0)$$

$$g_m = (0,x_m,0,0,0,0,0,0)$$

(where $l \neq m$)

as single-entry vectors (or scaled standard-basis vectors) representing a specific configuration of undulator gaps where all the IDs are open to the maximum and only the one identified with the location of the non-zero value moved to the gap size $x_{l}$ and $x_{m}$ respectively.  The system shows linearity when

$$ F_j\left( \sum_{i=1}^{N<N_{Gaps}} g_i \right)  =  \sum_{i=1}^{N<N_{gaps}}F_j\left(  g_i\right) $$
where with $F_j\left( \sum_{i=1}^{N<N_{Gaps}} g_i \right)$ we indicate the orbit distortion induced by the undulator gaps configuration obtained by summing the single-entry vectors.
This should hold for all possible gap combinations $\forall \{g_i\}$.

Hence, the criterion of linearity is:

$$ \Delta Prediction = || F_j\left( \sum_{1}^{N_{gaps}} g_i \right)  -  \sum_{1}^{N_{gaps}}F_j\left(  g_i\right) || \neq 0 $$

\begin{figure}[!h]
    \centering
    \includegraphics[width=\columnwidth]{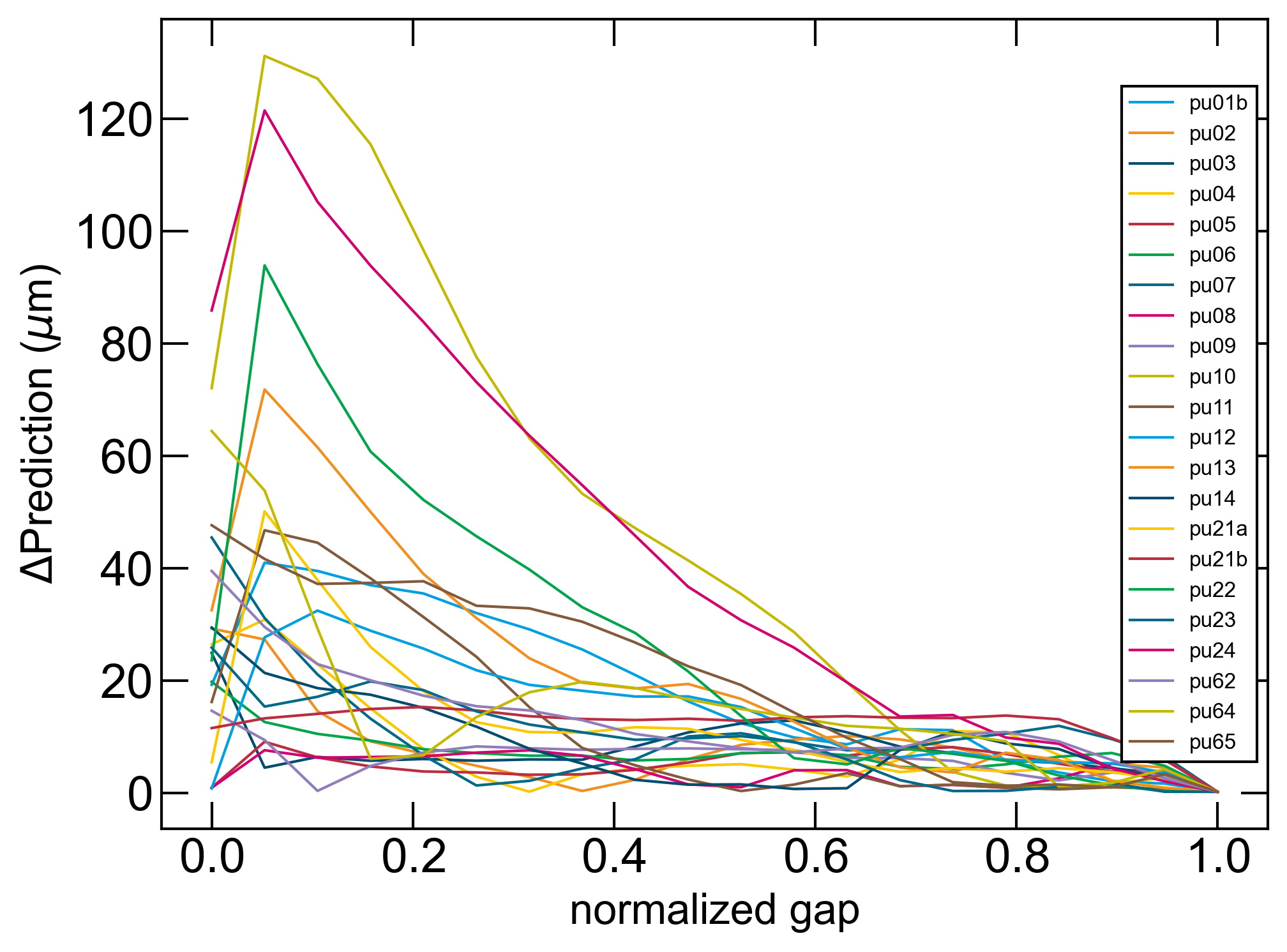}
    \caption{Difference of the predicted orbit (evaluated at one BPM) as calculated with the NN with two closed IDs (one of which is always PU01a) and as the sum of the predictions obtained for closing only one ID at a time. The gap of the undulators is linearly re-scaled to the 0-1 interval instead of the ID dependent min-max gap size.}
    \label{fig:lin_test}
\end{figure}

We performed the test by randomly assigning a gap value $g_0$ to PU01a and varying the gap in PU01b, while having all the other undulators open, and using the model to calculate the expected impact on the orbit. The operation was then repeated for all the other IDs. Figure~\ref{fig:lin_test} shows the difference between the predicted orbit distortions at one point along the ring for combinations of PU01a and any other undulator gap closing. The difference decreases to 0 when the undulator is open to the maximum gap as expected. 
Similar curves where obtained for all combinations of two undulators simultaneously closed. This indicates that the linearity criterion is not satisfied, suggesting that the impact of two or more undulators combined is not linear.

\section{CONCLUSIONS AND OUTLOOK}
Four NNs models were developed and trained with PETRA~III measurements to successfully predict the beam orbit at any given operational ID gaps configuration. The models were optimized through hyperparameter sweeps to obtain high predictability and comparisons indicated the 1D recurrent neural network as the best fitting. The NN trained on data sets where only one ID at a time was moving failed to reproduce the operational configurations where multiple gaps are varied simultaneously. 
Successful modeling was finally achieved by training a deep FFNN with a data set including multiple simultaneous gap variations.
The updated deep FFNN model reached a training loss of \SI{0.54}{\um} and a validation loss of \SI{0.46}{\um} after 200 epochs and showed good agreement with test data. To make sure the model was reliable we compared the predictions with measurements taken on different days.

The orbit distortion predicted by the trained model can then be used to calculate the required strength in the corrector magnets. The global correction scheme that we suggest uses a pyAT \cite{pyat} calculated orbit response matrix (ORM) with singular value decomposition \cite{orm}. Through the ORM it is possible to compute the kick at each of the correctors along the ring necessary to counteract the existing dipolar field errors such that the orbit distortion measured with the BPMs is minimized.

\begin{figure}[!tbh]
    \centering
    \includegraphics[width=0.35\textwidth]{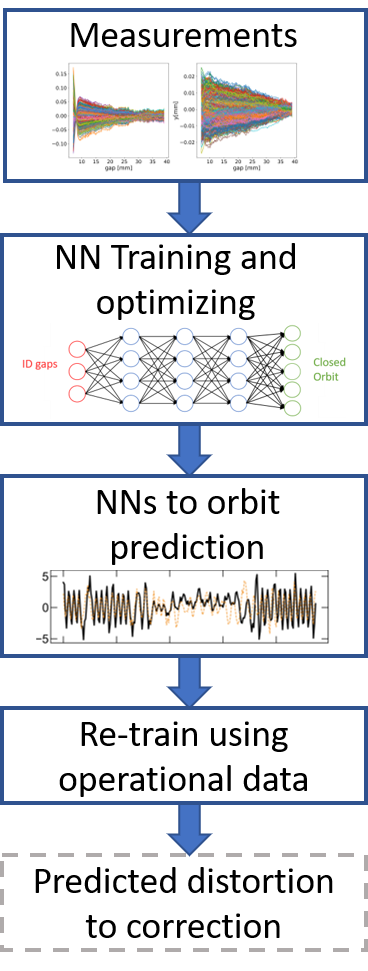}
    \caption{Description of the steps followed in this study to develop a reliable model to predict the undulator induced orbit distortion.}
    \label{fig:diagram}
\end{figure}

A diagram depicting the general process towards the orbit distortion correction is shown in Fig. \ref{fig:diagram}.

Although it has not yet been possible to evaluate an improvement in stability compared to the use of lookup tables, the advantage of developing such NN is to have a comprehensive and reliable model of the accelerator and paving the way for further advances. 
For example, a similar approach could also be considered to counteract the perturbation introduced by ID gap variations to the betatron coupling and the vertical dispersion, as demonstrated at ALS \cite{leeman}. The effect is currently not observable for PETRA~III specifications but could impact the extremely low emittances that will be reached in PETRA~IV \cite{petraiv}. In such case the impact (on the beam size) could be measured even with the fast orbit feedback activated thus collecting the necessary data during operations would be possible. In this way, the NN would be constantly updated, keeping up with drifts in the machine as well as changes in IDs applied by users without requiring additional dedicated measurement shifts. 

Moreover, the tools and methods developed here could potentially be adapted and applied to other accelerators, providing a versatile framework for improving beam stability across different machine configurations.

\section{ACKNOWLEDGEMENTS}
The authors thank Andreas Schoeps and his colleagues from the Undulator Group for the support for the measurements in PETRA III.

\clearpage
\newpage


\end{document}